\documentclass[a4paper,11pt]{article}
\usepackage{pos}

\title{Recent studies on top quark properties and mass in CMS}

\author*[1]{Dennis Schwarz}

\affiliation{Austrian Acedemy Of Sciences,\\
  Nikolsdorfer Gasse 18, 1050 Vienna, Austria}

\note{on behalf of the CMS Collaboration}

\emailAdd{dennis.schwarz@cern.ch}

\abstract{Studies of top quark properties using data collected by the CMS experiment are presented, including direct measurements of properties or extractions using differential cross section measurements. The latest results on top quark mass measurements using multiple kinematic distributions in a likelihood technique as well as the top quark pole mass derived from $\textrm{t}\overline{\textrm{t}}$+jet cross section measurements will be discussed.}

\FullConference{%
  41st International Conference on High Energy physics - ICHEP2022\\
  6-13 July, 2022\\
  Bologna, Italy
}

\begin{document}
\maketitle

\section{Introduction}
With a mass of about $172\;\text{GeV}$ and a Yukawa coupling close to unity, the top quark plays a special role in the electroweak sector of the standard model~(SM). Physics beyond the SM~(BSM) coupling to top quarks could manifest in subtle deviations from the SM predictions of top quark properties that can be probed with precise measurements at particle colliders. Due to the large center-of-mass energy and subsequently large production cross section of top quark final states, the Large Hadron Collider~(LHC) at CERN is an optimal laboratory for top quark physics. In this article, a selection of recent measurements by the CMS Collaboration of top quark properties and especially its mass $m_\text{t}$ is presented.  
A detailed description of the CMS experiment can be found in Ref.~\cite{CMSdetector}.

\section{Direct top quark mass measurement}
\begin{figure}
	\centering
	\includegraphics[width=.49\textwidth]{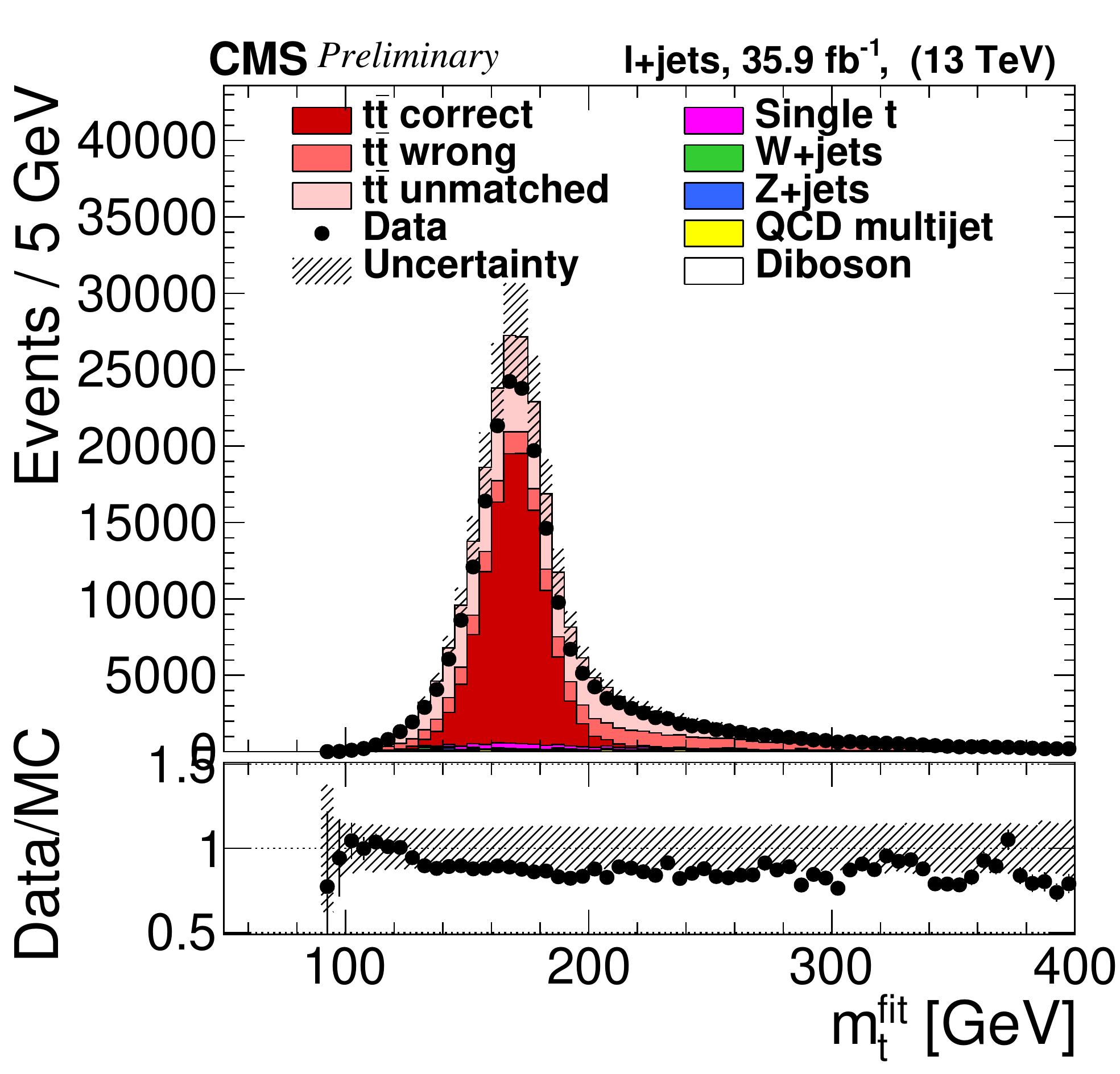}
	\caption{Fitted top quark mass $m_\textrm{t}^\textrm{fit}$. Simulated samples~(colored areas) are compared to data~(markers). The total uncertainty is displayed as hatched area and statistical uncertainties are indicated as vertical lines on the markers. The simulated $\textrm{t}\overline{\textrm{t}}$ events are split into contributions where the correct assignment of jets was found (correct), the jets were reconstructed but wrongly assigned (wrong) and the contribution where the reconstruction did not result in the correct set of jets (unmatched). Published in Ref.~\cite{direct}.}
	\label{f:direct}
\end{figure}

Very recently, the CMS Collaboration published the most precise direct measurement of the top quark mass to date~\cite{direct}. In the analysis, the lepton+jets decay channel of pair-produced top quarks ($\textrm{t}\overline{\textrm{t}}$) is selected. In a kinematic fit the $\textrm{t}\overline{\textrm{t}}$ system is reconstructed and a total of five independent observables are measured. The fitted mass $m_\textrm{t}^\textrm{fit}$ of the hadronically decaying top quark, shown in Fig.~\ref{f:direct}, is reconstructed from three separate jets and offers high sensitivity to the top quark mass. In addition, the ratio of the invariant mass of the lepton-b-jet system $m_\textrm{lb}^\textrm{fit}$ from the leptonically decaying top quark and $m_\textrm{t}^\textrm{fit}$ is used. This observable adds further information on the top quark mass from the leptonic leg of $\textrm{t}\overline{\textrm{t}}$ and is decorrelated from $m_\textrm{t}^\textrm{fit}$ by constructing the ratio. If the kinematic fit returns a low goodness-of-fit probability, $m_\textrm{t}^\textrm{fit}$ is not used and the leptonic observable $m_\textrm{lb}^\textrm{fit}$ is used directly. Also, the intermediate hadronically decaying W boson is reconstructed and its fitted mass $m_\textrm{W}^\textrm{fit}$ is measured. Finally, the momentum ratio of jets identified as b jet and light jets $R=(p_\textrm{T, b1}+p_\textrm{T, b2})/(p_\textrm{T, q1}+p_\textrm{T, q2})$ concludes the set of observables. While $m_\textrm{t}^\textrm{fit}$ and $m_\textrm{lb}^\textrm{fit}$ are sensitive to the top quark mass itself, $m_\textrm{W}^\textrm{fit}$ and $R$ can constrain the jet energy scale of light jets and b jets, respectively. These five observables are fit to data using a likelihood function, where uncertainties are profiled using nuisance parameters. By including all five observables, the leading uncertainties, mostly connected to the jet energy scale, can be largely constrained in the fit. Finally, the top quark mass is extracted by comparing data to simulated templates. Resulting in a value of $m_\text{t} = 171.77 \pm 0.38\;\text{GeV}$, this measurement provides the most precise determination of the top quark mass to date. 

\section{Extracting the top quark mass from cross sections}
\begin{figure}
	\centering
	\includegraphics[width=.49\textwidth]{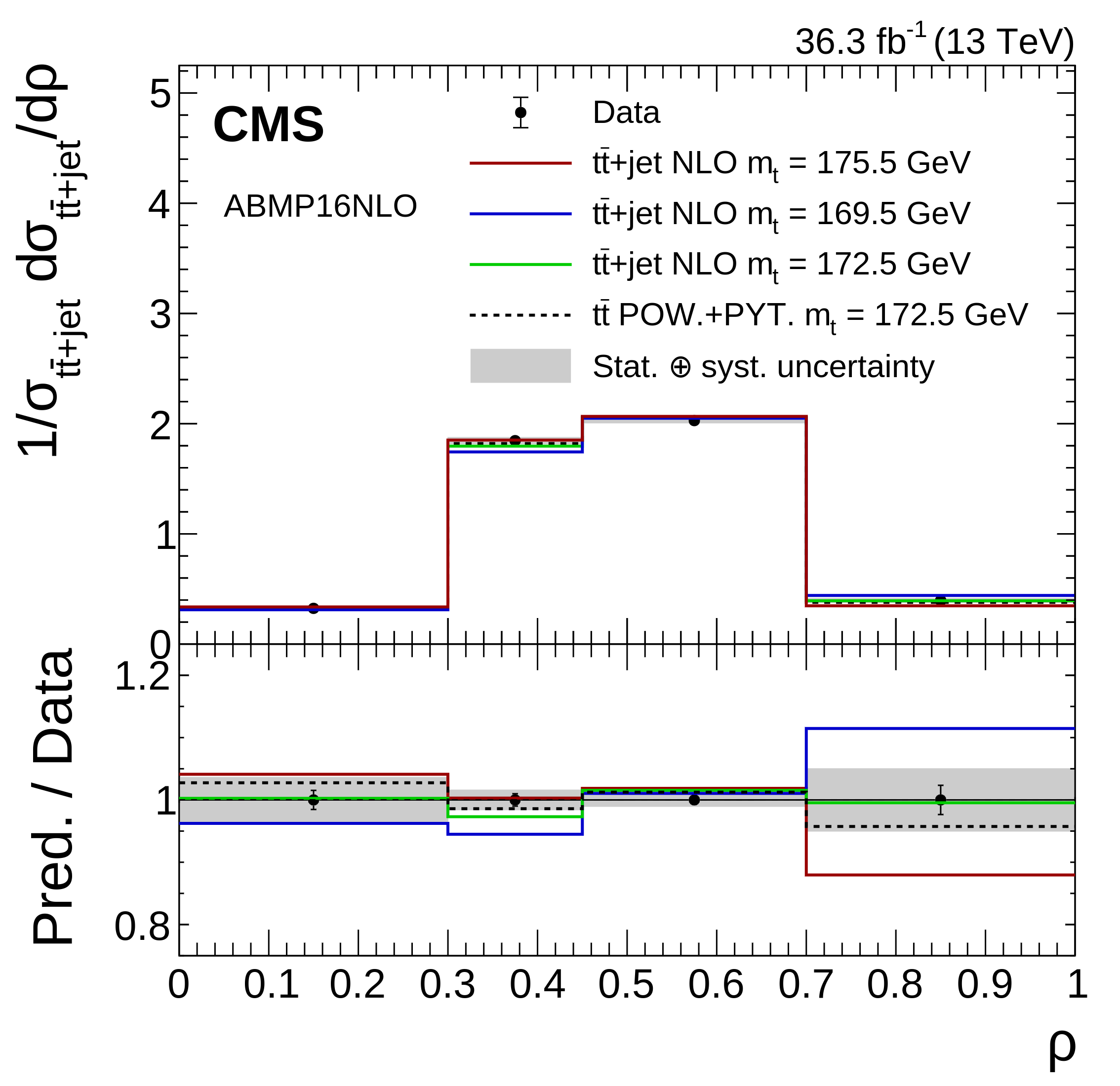}
	\includegraphics[width=.49\textwidth]{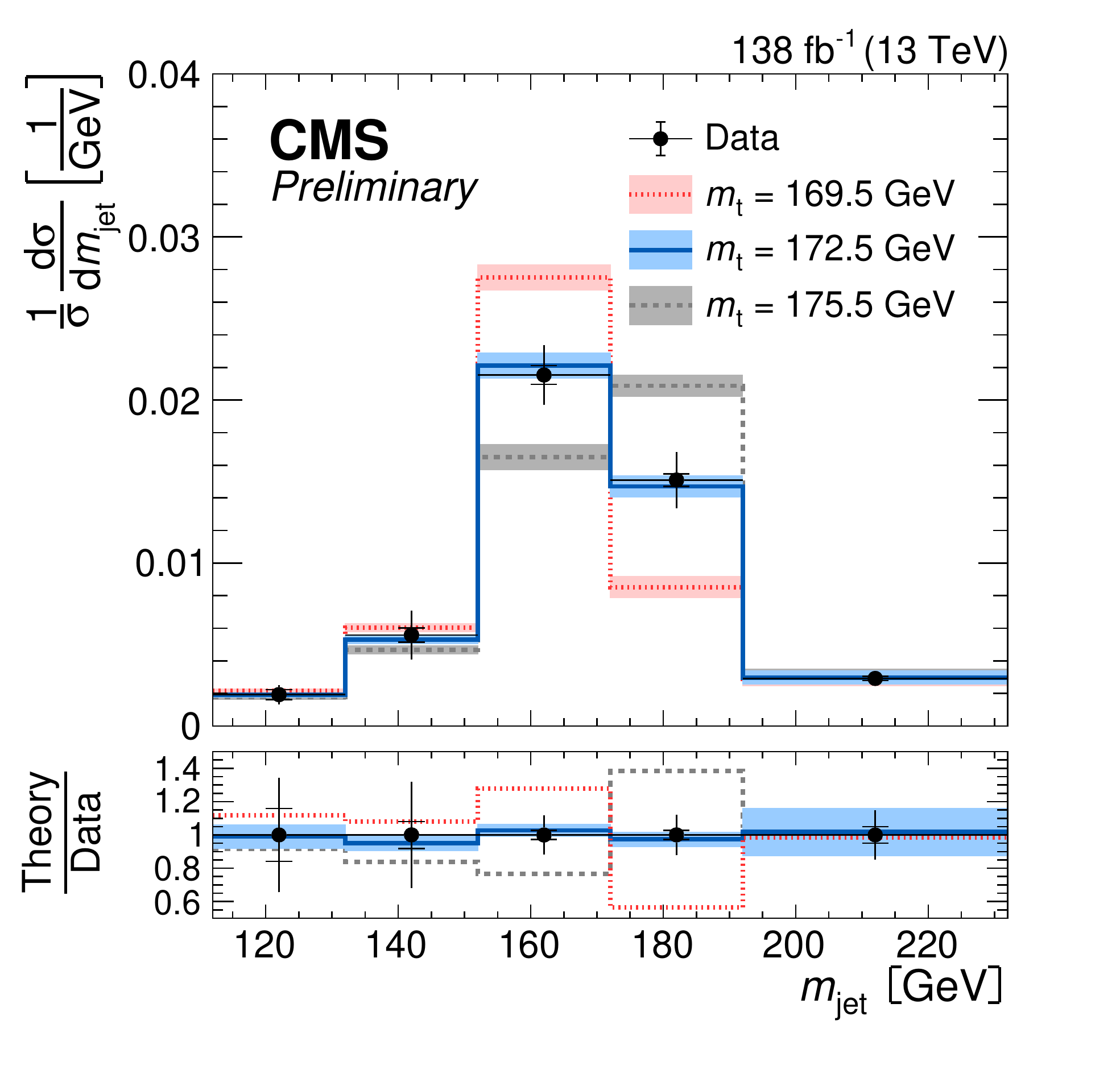}
	\caption{Normalized differential cross section as a function of $\rho$~(left). Data~(markers) are compared to analytical calculations of various top quark mass hypotheses. Normalized unfolded distribution in the jet mass~(right). Data~(markers) are compared to predictions for different values of the top quark mass obtained with simulation~(lines). The total uncertainties are shown with the vertical bars, the statistical component is indicated with the horizontal bars. Theoretical uncertainties are shown as colored areas. Published in Refs.~\cite{ttjet} and \cite{mjet}.}
	\label{f:massxs}
\end{figure}

Although direct measurements provide the most precise determinations of the top quark mass, there exist ambiguities in the definition of the top quark mass parameter in simulation~\cite{Hoang:2020iah}. Especially the parton shower modeling in simulation introduces non-perturbative QCD effects that are not analytically calculable. Thus, the relation to a well-defined mass scheme~\cite{Hoang:2017suc} introduces large theoretical uncertainties. These can be avoided by measuring inclusive or differential cross sections at the particle level that are sensitive to $m_\text{t}$ and can be accessed in calculations from first principles. This has been achieved in a measurement of the differential $\textrm{t}\overline{\textrm{t}}$ production cross section in association with one additional jet~\cite{ttjet}. Here, the dilepton channel of $\textrm{t}\overline{\textrm{t}}$ is selected and the cross section is measured as a function of $\rho=2m_0 / m_{\textrm{t}\overline{\textrm{t}}+\textrm{jet}}$, where $m_{\textrm{t}\overline{\textrm{t}}+\textrm{jet}}$ is the invariant mass of the fully reconstructed $\textrm{t}\overline{\textrm{t}}$+jet system and $m_0 = 170\;\text{GeV}$ is a constant. The reconstruction of $\rho$ is improved by using a neural network. A second neural network is trained to distinguish the signal from background processes. The unfolding is implemented within a profile likelihood fit, which makes it possible to constrain systematic uncertainties. The differential cross section is shown in Fig.~\ref{f:massxs}~(left). The kinematic region most sensitive to the value of the top quark mass is at the top quark pair production threshold at large values of $\rho$. The extraction of $m_\text{t}$ is performed by comparing the unfolded data to analytic predictions and results in values of $m_\text{t}^\text{pole} = 172.94 \pm 1.27\;(\text{fit+PDF+extr}) {}^{+0.51}_{-0.43}\;(\text{scale})\;\text{GeV}$ and $m_\text{t}^\text{pole} = 172.16 \pm 1.35\;(\text{fit+PDF+extr}) {}^{+0.50}_{-0.40}\;(\text{scale})\;\text{GeV}$, depending on the PDF set used. With the likelihood-based unfolding approach, the uncertainties in this measurement are drastically constrained, which results in an extraction of a well-defined top quark mass with a precision approaching $1\;\text{GeV}$.

So far, the most precise measurements of the top quark mass are driven by top quarks at low to medium energies. At high top-quark energies - due to the Lorentz-boost - the hadronic decay products of the top quark merge and can be reconstructed in a single large-radius jet. Then, jet substructure~\cite{Kogler:2021kkw,Asquith:2018igt} observables can be used to extract top quark properties. Especially the jet mass $m_\text{jet}$ - the invariant mass of the sum of the four-momenta of all jet constituents - shows sensitivity to the top quark mass. The CMS Collaboration drastically improved the precision of an earlier measurement~\cite{Sirunyan:2019rfa} by calibrating the jet mass scale and final state radiation modeling in the $\textrm{t}\overline{\textrm{t}}$ simulation. The measurement~\cite{mjet} is based on a customized two-step jet clustering with the XCone algorithm~\cite{Thaler:2015xaa, Sirunyan:2019rfa}. At first, two large-radius jets are found that aim at reconstructing the two top quark decays in an event. As a second step, three small-radius subjets within the large jets are clustered that target the three-prong top quark decay. Only particles that are clustered into one of the subjets are considered in the measurement. It is verified that even the visible two-prong structure of the leptonic top quark decay is well reconstructed using three subjets. The jet mass scale is calibrated by fitting the reconstructed W boson mass, that is obtained by selecting two out of the three XCone subjets. Here b-tagging information is used to identify the b jet in the top quark decay and subsequently selecting the other two subjets that reconstruct the W boson. The final state radiation modeling in the $\textrm{t}\overline{\textrm{t}}$ simulation is calibrated using jet substructure. The N-subjettiness~\cite{Thaler:2010tr} ratio $\tau_3/\tau_2$ in sensitive to the amount of final state radiation inside the measured jet that dilutes the expected three-prong decay. In this analysis, the simulation is calibrated to match the $\tau_3/\tau_2$ distribution observed in data. It is verified that no correlations to jet substructure observables exist that would affect the measurement. With these calibrations in place, the jet mass is unfolded to the particle level and the top quark mass is extracted from the normalized unfolded distribution, which is displayed in Fig.~\ref{f:massxs}~(right). So far, there are no analytic calculations available in the same phase space. Thus, the top quark mass is extracted from simulated templates and is measured to be $m_\text{t} = 172.76\pm0.81\;\text{GeV}$. This implies a gain in precision by about a factor of three compared to an earlier measurement~\cite{Sirunyan:2019rfa}. 
Approaching the precision of direct measurements at low top quark momenta, this analysis proves the capability of precision measurements of SM parameters in the high-energy regime. In addition, it is an important step in further understanding jet substructure observables that become more and more important in current LHC analyses.

\section{Charge asymmetry in top quark pair production}
\begin{figure}
	\centering
	\includegraphics[width=.49\textwidth]{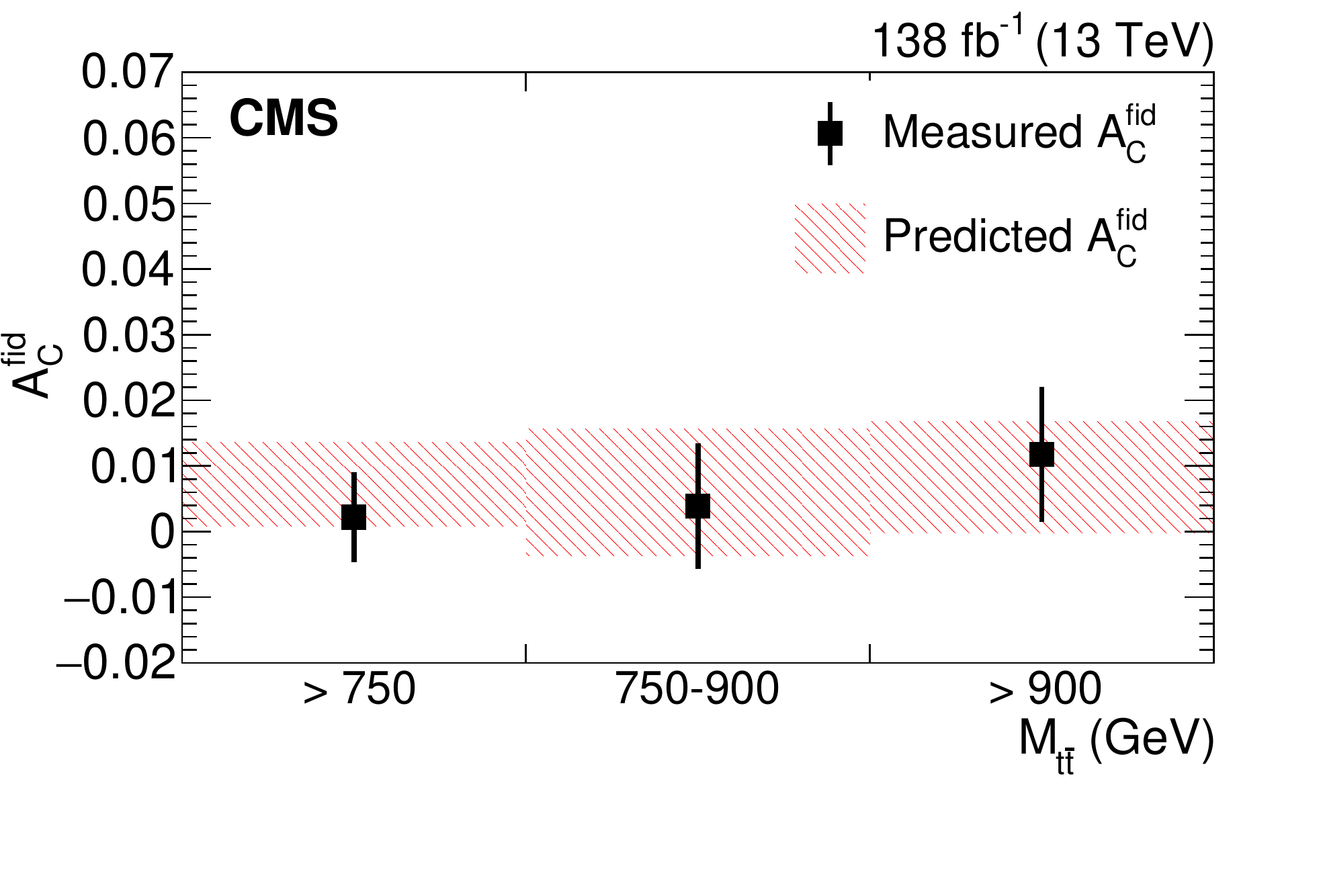}
	\includegraphics[width=.49\textwidth]{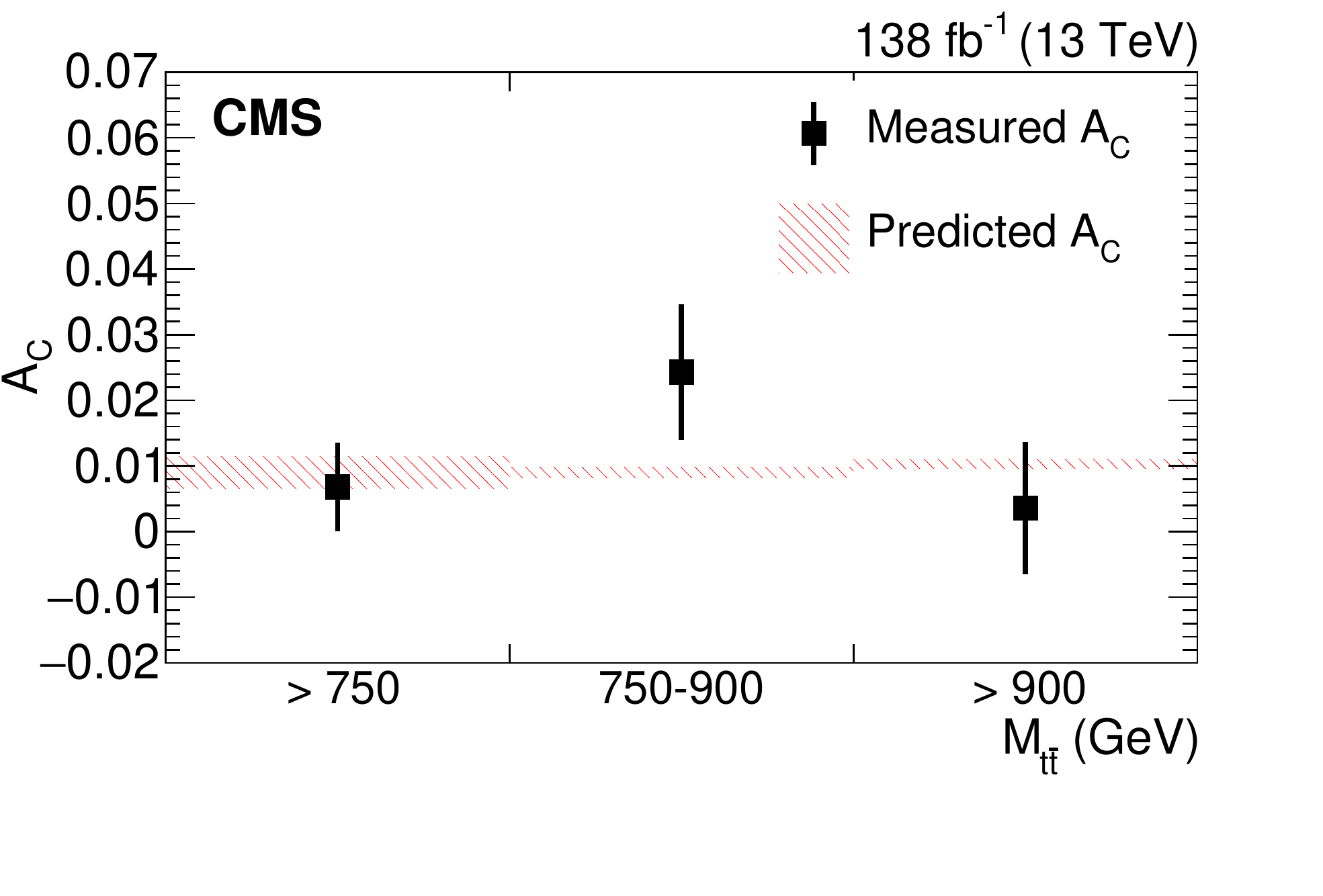}
	\caption{Charge asymmetry measured in the fiducial~(left) and full~(right) phase space in bins of the invariant mass of the $\textrm{t}\overline{\textrm{t}}$ system. Data~(markers) with total uncertainties are compared to the prediction of the SM~(red boxes). Published in Ref.~\cite{asym}.}
	\label{f:ac}
\end{figure}

In addition to the extensive program to measure the top quark mass in CMS, other properties of the top quark are studied as well. The measurement of the charge asymmetry in top quark final states in performed in Ref.~\cite{asym}. Here, the central-forward asymmetry of top quarks and top antiquarks is studied. In the SM at leading order, the asymmetry is predicted to be zero. At higher order however, there is a contribution in $\textrm{q}\overline{\textrm{q}} \rightarrow \textrm{t}\overline{\textrm{t}}$ production that leads to an asymmetry of about $1\%$. At the LHC, the dominant production mode of $\textrm{t}\overline{\textrm{t}}$ is via gluon-gluon fusion but $\textrm{q}\overline{\textrm{q}} \rightarrow \textrm{t}\overline{\textrm{t}}$ production is enhanced at high top quark energies because of the parton distribution functions of the colliding protons. Thus, the lepton+jets channel of $\textrm{t}\overline{\textrm{t}}$ is selected, where the invariant mass of the $\textrm{t}\overline{\textrm{t}}$ system surpasses $750\;\text{GeV}$. The hadronic top quark decay is reconstructed from either a combination of small-radius jets, a W-tagged large radius jet and a small-radius jet or a single top-tagged large-radius jet. The best reconstruction method is chosen by a $\chi^2$ minimization that reconstructs the full system. The asymmetry is then defined as
\begin{equation}
	A_C = \frac{N(\Delta|y|>0)-N(\Delta|y|<0)}{N(\Delta|y|>0)+N(\Delta|y|<0)},
\end{equation}
where $N(\Delta|y|>0)$ and $N(\Delta|y|<0)$ are the number of observed events, where the rapidity difference of top quark and top antiquark $(\Delta|y| = |y_t| - |y_{\bar{t}}|)$ are positive and negative, respectively. The data are unfolded with a maximum-likelihood approach, which allows to constrain systematic uncertainties, especially those connected to the background contribution. The acceptance of the event selection and reconstruction is estimated using simulated generator level information and the asymmetry is not only measured in the fiducial phase space but also obtained for the full phase space. As displayed in Fig.~\ref{f:ac}, the measurement is consistent with the SM prediction.

\section{Summary}
In this article, recent measurements of top quark properties performed by the CMS Collaboration were discussed. While there is an extensive effort to pin down the top quark mass using various methods and a variety of phase space regions, also other properties of the top quark are targeted. In many of these analyses, dedicated techniques were developed to reduce experimental systematic uncertainties and many measurements in the top quark sector are already limited by theoretical uncertainties.

\bibliographystyle{JHEP}
\bibliography{proceedings}

\end{document}